# High temperature oxygen NEXAFS valence band spectra and conductivity of LaFe$_{3/4}$Ni$_{1/4}$O$_3$ from 300 K to 773 K

Artur Braun[1,a], Selma Erat[1,2], Ahmad K. Ariffin[3,4]

Recardo Manzke[3], Hiroki Wadati[5], Thomas Graule[1,6], Ludwig J. Gauckler[2]

[1]Laboratory for High Performance Ceramics
Empa. Swiss Federal Laboratories for Materials Science and Technology
CH-8600 Dübendorf, Switzerland

[2]Department for Materials, Nonmetallic Inorganic Materials
ETH Zürich-Swiss Federal Institute of Technology, CH-8037 Zürich, Switzerland

[3]Institut für Physik, Humboldt-Universität zu Berlin
D-12489 Berlin, Germany

[4]Jabatan Fizik, Universiti Pendidikan Sultan Idris
35900 Perak, Malaysia

[5]Department of Applied Physics and Quantum-Phase Electronics Center (QPEC)

University of Tokyo, Tokyo 113-0032, Japan

[6]Technische Universität Bergakademie Freiberg
D-09596 Freiberg, Germany

---

[a] Corresponding author. Phone +41 (0) 58 765 4850, Fax +41 (0) 58 765 4150, email:

artur.braun@alumni.ethz.ch

**Abstract**

$LaFe_{3/4}Ni_{1/4}O_3$ was subjected to oxygen near edge x-ray absorption fine structure (NEXAFS) spectroscopy for 300 K < T < 773 K. The spectra show in the pre-edge a small hole doped peak originating from Ni substitution. The relative spectral weight of this transition to the weight of the hybridized O(2p) - Fe(3d) transitions scales with T and has a maximum at around 600 K. The characteristic energies of the thermal activated spectral intensity and conductivity suggest that the concentration of charge transferred electrons from O(2p) to Ni(3d) increases and that the pre-edges account in part for the polaron activated transport.

**Introduction**

The electronic structure of materials relevant for solid oxide fuel cell cathodes such as $La_{1-x}Sr_xMeO_3$ (Me=Mn, Fe, Co, Ni) has been subject of increasing interest in the last decades [1]. Electronic structure studies are limited to ambient temperature in the context of fuel cell materials, and even lower temperatures in the context of condensed matter theory. There are numerous high temperature (T → 1200 K) transport studies and studies which address the crystallographic structure and their changes as a function of stoichiometry, temperature, and oxygen partial pressure, in particular high temperature neutron diffraction and x-ray diffraction studies. Interestingly, there are virtually no studies on the electronic structure of cathode materials for the high operation temperatures.

We have recently reported valence band photoemission spectroscopy (VB PES) studies on LaSrFe-based and LaSrFeNi-based perovskites in ultra-high vacuum at temperatures up to 673 K and 831 K, respectively. In these studies could we find a correlation of details in the spectral characteristic of the VB PES data and in the transport characteristics, suggesting that VB PES is an adequate method to study the transport properties of ceramics at high temperatures at the molecular level [2-4].

Because the valence band and the density of states near the Fermi level are the relevant parameter for the functionalities of materials, we employed also near edge x-ray absorption fine structure (NEXAFS) spectroscopy at the oxygen K-edge. Encouraged by the finding that the ratio of spectral weight of electron hole states and hybridized states in the pre-edges of the oxygen NEXAFS spectra of LaSrFe-based materials scale quantitatively with their electronic conductivity [5], we performed oxygen NEXAFS spectroscopy for LaFeNi-oxide at temperatures from 300 K to 773

K. Here we found that the intensity of the Ni derived states increase with increasing temperature.

From an Arrhenius plot of the conductivity data and spectral intensity we were able to derive activation energies, which are 151.7 meV ± 1.3 meV for the conductivity and 66 meV ± 34.5 meV for the change of the spectral weight.

In the spirit of a recent presentation of the temperature dependency of VB PES and NEXAFS spectra data of LaSrFe-oxide between low temperature and 300 K in a schematic sketch of the density of states (DOS) [6], we combine the aforementioned high temperature VB PES data of LaSrFeNi-oxide and LaSrFe-oxide, and the high temperature NEXAFS data of the present study, in order to sketch changes of the DOS at the Fermi energy from 300 K to 773 K.

**Experimental**

Polycrystalline $LaFe_{0.75}Ni_{0.25}O_{3-\delta}$ was prepared by conventional solid state reaction. The precursors $La_2O_3$ (>99.99 %), $SrCO_3$ (99.9 %), $Fe_2O_3$ (>99.0 %) and NiO (99.8 %) were mixed in stoichiometric proportions, calcined at 1473 K for 4 h and then sintered at 1673 K for 12 h with 5 K/min heating/cooling rates. The powders were pressed into 5 mm x 3 mm x 25 mm bars and sintered at 1673 K for 12 h with the same heating/cooling rate [7].

The crystallographic structure has been determined by x-ray diffraction at 300 K: $LaFe_{0.75}Ni_{0.25}O_3$) has orthorhombic symmetry with space group *Pbnm* (62),

The 4-point DC conductivity of the samples was measured in a furnace in air at a temperature range of 300 K to 1273 K [7].

Temperature dependent NEXAFS spectra were recorded at UE56/2-PGM1-beamline at BESSY II using the BESSY/HU chamber, with the samples mounted on aluminum sample holders and measured in vacuum with base pressure of about $10^{-8}$ mbar. Signal detection was carried out in the total electron yield (TEY) mode. Temperature was controlled by resistive heating of the sample holder and passive cooling, and measured with a thermocouple in direct contact with the sample.

Background subtraction and normalization were performed by subtracting linear least square fits for x-ray energies < 525 eV and > 550 eV. The relative uncertainties in the peak height ratios, as determined by deconvolution of the spectra into Voigt functions served as input for error propagation in order to obtain the uncertainty for the relative spectral weight S (peak height ratio). Data reduction and analysis was performed with WinXAS [9].

**Results and Discussion**

The crystallographic and electronic structure and the near edge x-ray absorption fine structure (NEXAFS) spectrum at the oxygen K-shell absorption edge of the charge transfer insulator $LaFeO_3$ are well understood [5,10-12]. Figure 1 shows the oxygen NEXAFS spectrum of $LaFeO_3$ as reproduced after Wu et al. [10] and for comparison our $LaFe_{3/4}Ni_{1/4}O_3$ spectrum with deconvolution into particular resonances with spectral assignment. $LaFeO_3$ has a characteristic doublet in the valence band with transitions from Fe3d-O2p hybridized states with spin up $t_{2g}$ and $e_g$ symmetry at around 529 and 530 eV. Substitution on the A-site with Sr for example forms electron holes which manifest in an extra transition with spin up $e_g$ symmetry just before this doublet [13]. Similar holds for B-site substitution with Ni, for example [14], which we notice upon close inspection of our spectrum of $LaFe_{3/4}Ni_{1/4}O_3$ in Figure 1 (bottom), where we see an extra small peak at about 527 eV which is absent in the spectrum of $LaFeO_3$ (top). The relative spectral weight $S = \frac{e_g\uparrow}{t_{2g}\downarrow + e_g\downarrow}$ of the *hole* peak and the doublet in LSF-based iron perovskites is quantitatively correlated with the electric conductivity; this is particularly evident in systematic substitution studies such as shown in [4] on Ta and Ti substituted LSF.

The oxygen NEXAFS spectra of $LaFe_{3/4}Ni_{1/4}O_3$ in Figure 2 were recorded at temperatures from 300 K to 773 K and precisely deconvoluted so as to be able to accurately determine the relative peak heights and thus the relative spectral weight S. The $e_g\uparrow$ peak at 300 K is barely visible and has an intensity of 0.08. The relative intensities of the $t_{2g}\downarrow$ and $e_g\downarrow$ peaks are 1.600 and 1.338, respectively. We then find that S ranges from 0.027 to 0.076 in the temperature range from 300 K to 773 K. The upper temperature range is relevant for intermediate temperature solid oxide fuel cells, for example [15].

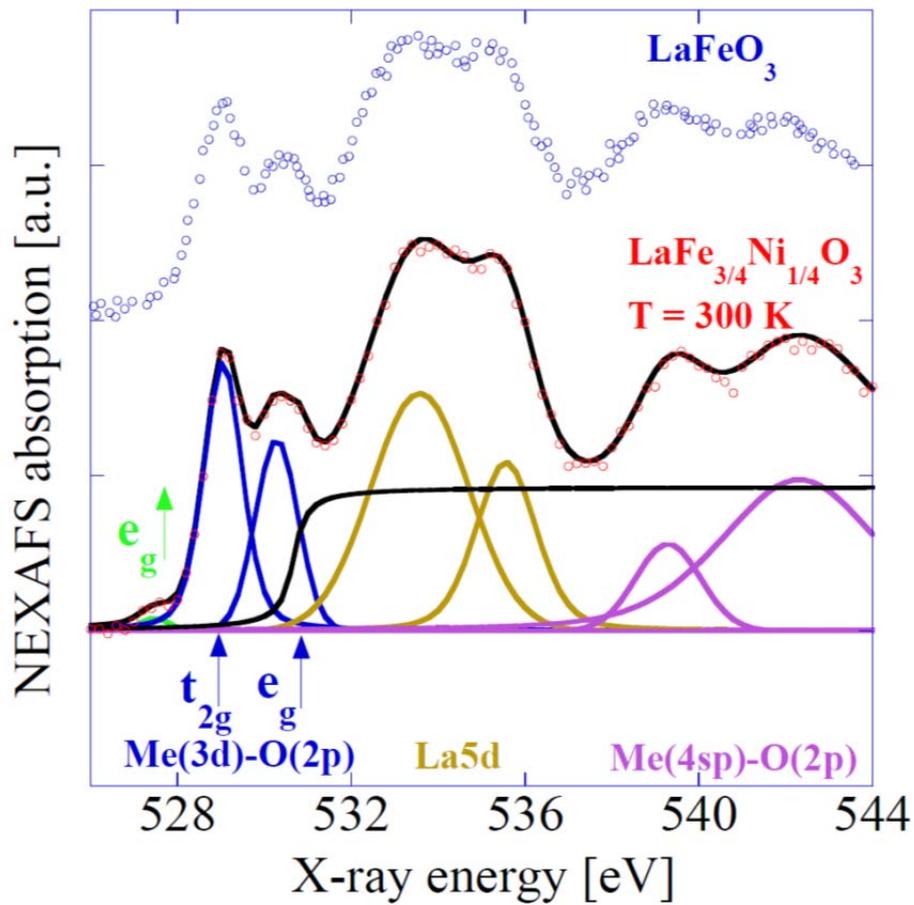

**Figure 1:** Oxygen 1s absorption spectrum of LaFe$_{3/4}$Ni$_{1/4}$O$_3$ recorded at 300 K (bottom; open symbols are data points, solid line is least square fit) in ultra-high vacuum with suggested peak assignment and deconvolution into Voigt functions, and spectrum of LaFeO3 (top, open symbols) after Wu et al. [11].

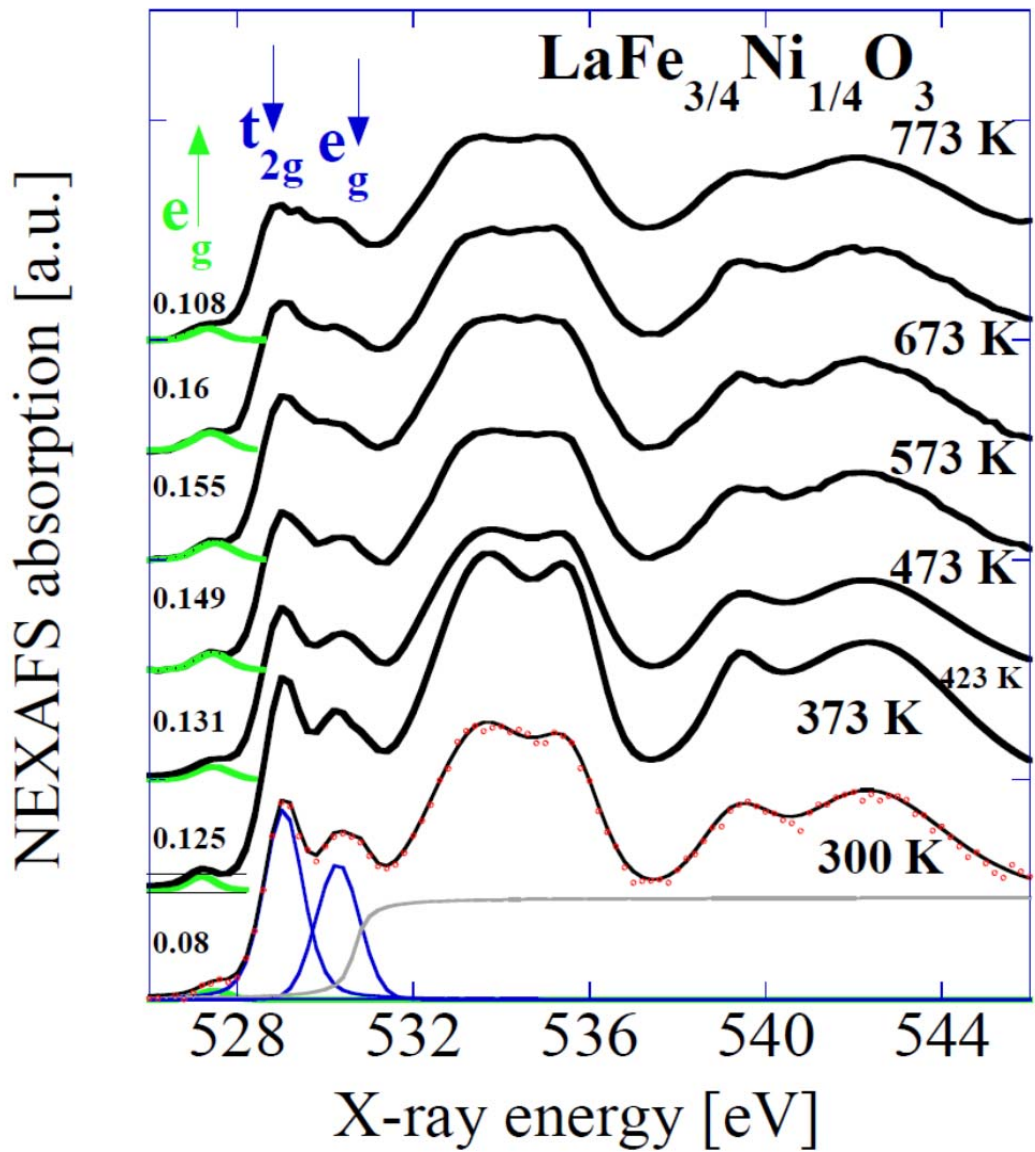

**Figure 2:** Oxygen 1s absorption spectra recorded at temperatures from 300 K to 773 K in ultra-high vacuum. Spectra are vertically offset by 1 unit for better comparison. Numbers on the left show the relative peak height of $e_g\downarrow$; numbers on the right denote the temperatures.

The electronic conductivity in LaSrFe-oxides [16] and LaSrFeNi-oxides and many oxides are explained in term of thermally activated small polaron hopping [17,18], which involves the trapping of an electron or electron hole at one crystallographic site by the local lattice polarization which it causes [19]. The temperature dependent conductivity σ(T) for small polaron hopping conductors is given by an exponential relationship of the form $\sigma \cdot T \propto \exp(E_a/k_B T)$ with activation Energy $E_a$ and Boltzmann's constant k [20-24]. Figure 3 (top) shows the Arrhenius plot of conductivity σ and relative spectral weight S for the sample. σ and S increase upon temperature increase in the intermediate temperature range. σ starts to decrease at T > 895 K whereas S starts to decrease at lower temperature around T > 600 K.

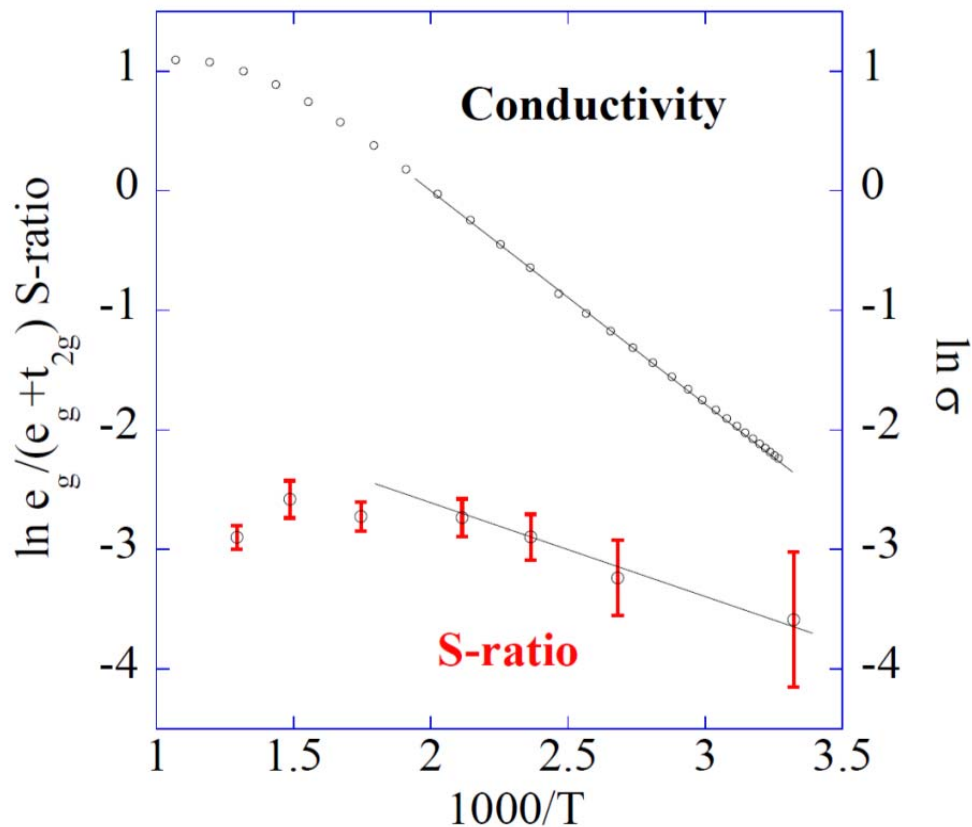

**Figure 3**: Arrhenius plot for the conductivity (top, right axis) and spectral ratio S (bottom, with error bars, left axis). Solid lines denote linear least square fits for the determination of the activation energy.

The activation energy was determined in the temperature range of 305 K ≤ T ≤ 555 K to around 151.7 meV.

LaFe$_{0.75}$Ni$_{0.25}$O$_3$ has distorted orthorhombic symmetry in which the super exchange angle <Fe/Ni–O–Fe/Ni> deviates from 180°. In addition it has completely Fe$^{3+}$ configuration 3d$^5$ at 300 K, where the hopping mechanism takes place in the super exchange unit Fe$^{3+}$–O–Fe$^{3+}$ with antiferromagnetic super exchange interaction [8]. Since the hopping mechanism occurs via O bridge, the oxygen vacancies cause a decrease in conductivity.

The symmetry of the lattice increases with temperature resulting in an increase in the super exchange angle. This enhances overlapping between Fe/Ni (3d) and O (2p) orbitals. The more overlapping of these orbitals decreases the electron-electron interaction which is one of the reasons for resistivity.

The spectral ratio S in which e$_g$↑ states are created due to hybridization between Ni (3d) and O (2p) and the e$_g$↓ + t$_{2g}$↓ states are due to hybridization between Fe(3d) and O (2p) increases with temperature up to 600 K, and then decreases. As we see in Figure 2, the Ni derived empty e$_g$↑ states are closer to E$_F$ than the two Fe derived states. The Ni derived states are hence the more favourable states for charge transfer mechanism which amounts to higher conductivity. This is the rational why LaFeO$_3$ is an insulator with 2 eV charge transfer band gap [5] whereas LaFe$_{0.75}$Ni$_{0.25}$O$_3$ is a semiconductor. Since the S ratio increases with increasing temperature, we conclude that the concentration of charged transferred electrons from O(2p) to Ni(3d) (creating an electron hole on O site) is increased by temperature up to 600 K. However, at higher temperature the concentration of charge transferred electrons is reduced most probably because of oxygen vacancies present. The temperature dependent changes in S-ratio and in conductivity are not really paralleled. Therefore, it is worth to mention that LaFe$_{0.75}$Ni$_{0.25}$O$_3$ has two different mechanisms which contribute to the conductivity: electron hopping and charge transfer.

Figure 4 shows the density of states for LaSrFeNi-oxide for temperatures from 300 K to 773 K as derived from VB PES [1,2] and oxygen NEXAFS. Occupied (in photoemission) and unoccupied states (in NEXAFS) are affected by temperature. Interestingly, Sr doped LaFeO$_3$ and Ni doped LaFeO$_3$ show the e$_g$↑ peak just above E$_F$ virtually at the same energy. In the former case (heterovalent substitution), since La$^{3+}$ is replaced by Sr which has lower valence (2+) the "*new empty states*" so called "*hole doped peak*" are created. In the latter case (homovalent substitution), Ni affects the unoccupied states and cause to redistribute these states and finally some states moved towards $E_F$. Further, these *redistributed states* increases up to 600 K.

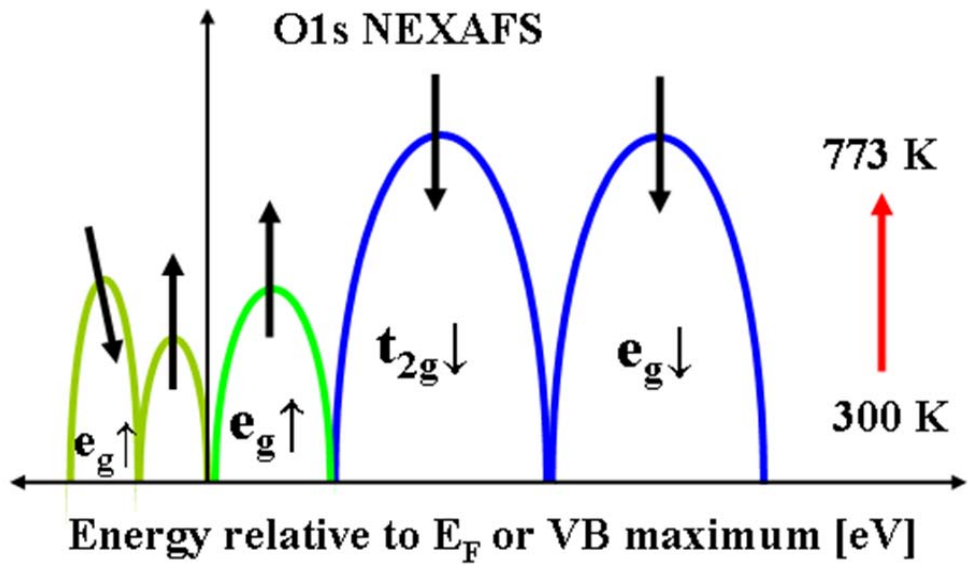

**Figure 4**. Schematic for the density of states for LaSrFeNi-oxide for temperatures from 300 K to 773 K as derived from VB PES [previous work in APL] and oxygen NEXAFS. The red arrow represents increasing temperature and. The direction of the arrow shows whether a structure increases/decreases in intensity, or shifts along the energy axis [6].

**Conclusion**

$LaFeO_3$ doped with Ni shows an extra pre peak due to Ni(3d)-O(2p) hybridization observed in O K edge NEXAFS spectra and the intensity of this peak increases upon temperature increase. In contrast to Sr doping, Ni doping does not create new hole states but cause redistribution of empty states. The electric conductivity of $LaFe_{0.75}Ni_{0.25}O_3$ is explained by two different thermally activated mechanisms, hopping process in $Fe^{3+}$-O-$Fe^{3+}$ chain and charge transfer process from O(2p) to mainly Ni(3d). The temperature dependent spectral weight and conductivity shows similar trend but not exactly the same which allows us to conclude that the differences is because of the hopping process. The decrease in spectral weight is at lower temperature than semiconducting to metallic like conductivity transition temperature. This can be attributed to the differences in measurement environment: the O K edge NEXAFS spectra were measured in UHV whereas conductivity was measured in open air. The differences in the measurement environment affect the oxygen vacancies. The vacancies are more easily created in vacuum than in air.

**Acknowledgements**

The research leading to these results received funding from the European Community's Sixth Framework Marie Curie International Reintegration Programme grant n°. 042095, Sixth Framework Programme (Real-SOFC n°. SES6-CT-2003-502612), and Swiss National Science Foundation grant n°. 200021-116688. We acknowledge Helmholtz-Zentrum Berlin - Electron storage ring BESSY II for provision of synchrotron radiation at beamline MUSTANG UE-56 and are indebted to A. Vollmer for supplying us with the high-temperature sample holder and W. Mahler, M. Sperling B. Zada, T. Blume for support during the NEXAFS experiments.